\documentclass[rmp,aps,preprint,nofootinbib,endfloats*,a4paper,notitlepage,groupedaddress, superscriptaddress]{revtex4}
\usepackage{amsmath}
\usepackage{amssymb}
\usepackage{amsfonts}
\usepackage{epsfig, color, ulem}
\usepackage{graphicx}

\usepackage{setspace}

\newlength{\dinwidth}
\newlength{\dinmargin}
\usepackage{lineno}
\newcommand{\bftau}{\mbox{\boldmath $\tau$}}

\begin{document}

\begin{spacing}{1.2}

\title{Reciprocity and representation theorems for rotational seismology}
\author{\small Kees Wapenaar \\
Department of Geoscience and Engineering, Delft University of Technology,  The Netherlands}

\date{\today}

\begin{abstract}
{\small Recently, there has been an increasing interest in employing rotational motion measurements for seismic source inversion, structural imaging and ambient noise analysis.
We derive reciprocity and representation theorems for rotational motion.
The representations express the rotational motion inside an inhomogeneous anisotropic earth in terms of translational and rotational motion at the surface.
The theorems contribute to the theoretical basis for rotational seismology methodology, such as determining the moment tensor of earthquake sources.}
\end{abstract}

\maketitle

\section{1\,\, Introduction}

Measurements of the seismic wave field are traditionally restricted to three mutually perpendicular components of the particle velocity (also called translational motion).
Observational studies by \citet{Igel2007GJI}, \citet{Lin2011GRL} and others have demonstrated the potential of additionally measuring three components of the rotational motion.
Recently, researchers have been exploring the advantages of rotational seismology in source localization and inversion 
\citep{Bernauer2014JGR, Donner2016GJI, Li2017GJI, Ichinose2021JGR}, structural imaging \citep{Bernauer2009GEO, Abreu2023GJI},
ambient noise analysis \citep{Hadziioannou2012JS, Paitz2019GJI}, and exploration geophysics \citep{Li2017GEO, Schmelzbach2018GEO}.

Reciprocity and representation theorems for translational motion \citep{Hoop66ASR, Aki80Book, Fokkema93Book}
have been employed as a theoretical basis for the development of methodologies for seismic imaging, inverse scattering,
source characterization, seismic holography, Green's function retrieval, etc.
Given the current interest in rotational seismology, it is opportune to derive reciprocity and representation theorems for rotational motion.
An important step in this direction has been made by \citet{Li2017GJI}. In their derivation they assume
the medium is homogeneous and isotropic. In this paper we derive several forms of reciprocity and representation theorems 
for translational and rotational motion in an inhomogeneous anisotropic earth.  
These theorems complement the theoretical basis for the development of methodologies for rotational seismology.

\section{2\,\, Reciprocity theorems for rotational seismology}\label{sec1}

The rotational motion-rate vector in an inhomogeneous anisotropic medium is defined as $\dot{\bf \Omega}=\frac{1}{2}{\bf\nabla}\times{\bf v}$,
where ${\bf v}$ is the particle velocity vector. This definition holds in the space-time $({\bf x},t)$ domain as well as in the space-frequency $({\bf x},\omega)$ domain.
In the following, all expressions are in the space-frequency domain.
Note that for the special case of a homogeneous isotropic medium, $\dot{\bf \Omega}$ would represent the $S$-wave part of ${\bf v}$.

Consider a spatial domain $\mathbb{D}$ enclosed by boundary $\partial\mathbb{D}$ with outward pointing normal vector ${\bf n}$. 
A reciprocity theorem for elastic wave fields in a homogeneous isotropic medium in this domain  reads
\begin{eqnarray}
&&\hspace{-.8cm}\oint_{\partial\mathbb{D}}\rho\Bigl[c_P^2\{{\bf v}_A{\bf\nabla}\cdot{\bf v}_B - {\bf v}_B{\bf\nabla}\cdot{\bf v}_A\}
+c_S^2\{{\bf v}_A\times{\bf\nabla}\times{\bf v}_B - {\bf v}_B\times{\bf\nabla}\times{\bf v}_A\}\Bigr]\cdot{\bf n}{\rm d}^2{\bf x}\nonumber\\
&&\hspace{.43cm}=i\omega\int_\mathbb{D}\{{\bf v}_A\cdot{\bf f}_B - {\bf v}_B\cdot{\bf f}_A\}{\rm d}^3{\bf x}.\label{eqS2}
\end{eqnarray}
Here  ${\bf v}({\bf x},\omega)$ is the particle velocity vector, ${\bf f}({\bf x},\omega)$ the  force source vector, $c_P$ and $c_S$ are the $P$- and $S$-propagation velocities, $\rho$ is the 
mass density and $i$ the imaginary unit. Upper-case subscripts $A$ and $B$ denote two independent states, which can be physical or mathematical wave fields 
(or a combination thereof), emitted by different sources. Equation (\ref{eqS2}) is a slightly modified form of a theorem formulated by \citet{Knopoff56JASA}. 
Because it explicitly contains ${\bf\nabla}\times{\bf v}=2\dot{\bf \Omega}$ in both states, \citet{Li2017GJI} used this as the starting point for deriving representations for rotational seismology.
A limitation is that equation (\ref{eqS2}) was derived from the elastic wave equation for a homogeneous isotropic medium. 
Here we show that its derivation can be generalized for an inhomogeneous anisotropic medium in $\mathbb{D}$, 
with only some restrictions on the medium parameters at the boundary $\partial\mathbb{D}$.

The Betti-Rayleigh reciprocity theorem for elastic wave fields in an inhomogeneous anisotropic medium in $\mathbb{D}$ reads \citep{Hoop66ASR, Aki80Book} 
\begin{eqnarray}
&&\oint_{\partial\mathbb{D}}\{v_{i,A}\tau_{ij,B}-v_{i,B}\tau_{ij,A}\}n_j{\rm d}^2{\bf x}=-\int_\mathbb{D}\{v_{i,A}f_{i,B}-v_{i,B}f_{i,A}\}{\rm d}^3{\bf x}.\label{eq9}
\end{eqnarray}
Here $v_i$, $\tau_{ij}$, $f_i$ and $n_j$ are components of the particle velocity vector ${\bf v}$, stress tensor $\bftau$, force source vector ${\bf f}$
and normal vector ${\bf n}$, respectively.  Einstein's summation convention applies to repeated lower-case subscripts. 
Our aim is to recast equation (\ref{eq9}) into the form of equation (\ref{eqS2}). 
We start by expressing the stress at the boundary $\partial\mathbb{D}$ in terms of the particle velocity. 
Because the boundary integral in equation (\ref{eqS2}) contains the isotropic velocities $c_P$ and $c_S$, for our derivation we assume that
 the medium is isotropic in a vanishingly thin shell around $\partial\mathbb{D}$. Hence, we use the isotropic stress-velocity relation
$\tau_{ij}=-\frac{1}{i\omega}\{\lambda\delta_{ij}\partial_kv_k + \mu(\partial_jv_i+\partial_iv_j)\}$,
where $\lambda({\bf x})$ and $\mu({\bf x})$ are the Lam\'e parameters in the thin shell around $\partial\mathbb{D}$.
Substituting this into equation (\ref{eq9}) for both states, we obtain after some manipulations
\begin{eqnarray}
&&\hspace{-.83cm}\oint_{\partial\mathbb{D}}\Bigl[\lambda\{{\bf v}_A{\bf\nabla}\cdot{\bf v}_B-{\bf v}_B{\bf\nabla}\cdot{\bf v}_A\}
+2\mu\{({\bf v}_A\cdot{\bf\nabla}){\bf v}_B - ({\bf v}_B\cdot{\bf\nabla}){\bf v}_A\}\nonumber\\
&&\hspace{.43cm}+\mu\{{\bf v}_A\times{\bf\nabla}\times{\bf v}_B - {\bf v}_B\times{\bf\nabla}\times{\bf v}_A\}\Bigr]\cdot{\bf n}{\rm d}^2{\bf x}
=i\omega\int_\mathbb{D}\{{\bf v}_A\cdot{\bf f}_B - {\bf v}_B\cdot{\bf f}_A\}{\rm d}^3{\bf x}.\label{eq14}
\end{eqnarray}
For a more detailed derivation see Appendix A. 
This expression has the form of equation (\ref{eqS2}), except for the second term on the left-hand side. 
This term can be reorganized into the form of the first term, using the theorem of Gauss, if we assume $\mu$ 
is constant along the boundary $\partial\mathbb{D}$ (see Appendix A for details). Using $\lambda+2\mu=\rho c_P^2$ and $\mu=\rho c_S^2$, we thus obtain
equation (\ref{eqS2}), this time for an inhomogeneous anisotropic medium in $\mathbb{D}$;
only in a vanishingly thin shell around the boundary $\partial\mathbb{D}$ the medium is assumed to be isotropic, with $\mu$ constant along $\partial\mathbb{D}$.

In order to use equation (\ref{eqS2}) as a basis for backpropagation, we replace the quantities in state $A$ by their complex-conjugates (denoted by asterisks), 
which is allowed when the medium in $\mathbb{D}$ is lossless. Using $\dot{\bf \Omega}=\frac{1}{2}{\bf\nabla}\times{\bf v}$ and defining the cubic dilatation-rate as $\dot\Theta={\bf\nabla}\cdot{\bf v}$
we thus obtain
\begin{eqnarray}
&&\hspace{-.8cm}\oint_{\partial\mathbb{D}}\rho\Bigl[c_P^2\{{\bf v}_A^*\dot\Theta_B-{\bf v}_B\dot\Theta_A^*\}
+2c_S^2\{{\bf v}_A^*\times\dot{\bf \Omega}_B - {\bf v}_B\times\dot{\bf \Omega}_A^*\}\Bigr]\cdot{\bf n}{\rm d}^2{\bf x}
=i\omega\int_\mathbb{D}\{{\bf v}_A^*\cdot{\bf f}_B + {\bf v}_B\cdot{\bf f}_A^*\}{\rm d}^3{\bf x}.\nonumber\\
&&\label{eq19b}
\end{eqnarray}

 \section{3\,\, Representation theorems for rotational seismology}

A representation theorem is obtained by choosing for one of the states in a reciprocity theorem a Green's state \citep{Gangi70JGR}. Hence,
for state $A$, we replace ${\bf v}_A$ by the Green's velocity vector ${\bf G}_{v,f_n}({\bf x},{\bf x}_A,\omega)$, defined as the response at ${\bf x}$ 
to a unit force source in the $x_n$-direction at ${\bf x}_A$ in $\mathbb{D}$, i.e., $f_{i,A}({\bf x},\omega)=\delta_{in}\delta({\bf x}-{\bf x}_A)$. 
Moreover, we replace  $\dot{\bf \Omega}_A$ and $\dot\Theta_A$ by the Green's  rotational motion-rate vector and cubic dilatation-rate, defined as 
${\bf G}_{\dot\Omega,f_n}({\bf x},{\bf x}_A,\omega)=\frac{1}{2}{\bf\nabla}\times{\bf G}_{v,f_n}({\bf x},{\bf x}_A,\omega)$ and
$G_{\dot\Theta,f_n}({\bf x},{\bf x}_A,\omega)={\bf\nabla}\cdot{\bf G}_{v,f_n}({\bf x},{\bf x}_A,\omega)$.
As usual, the second subscript of a Green's function refers to the source type at ${\bf x}_A$ 
(here a force $f_n$), whereas the first subscript ($v$,  $\dot\Omega$ or $\dot\Theta$)
refers to the type of response at ${\bf x}$. For state $B$ we take the actual physical state and drop the subscripts $B$. 
Choosing the source distribution ${\bf f}({\bf x},\omega)$ of the actual state outside $\mathbb{D}$, we thus obtain from equation (\ref{eq19b})
\begin{eqnarray}
v_n({\bf x}_A,\omega)=\frac{1}{i\omega}\oint_{\partial\mathbb{D}}\rho\Bigl[c_P^2\{{\bf G}_{v,f_n}^*\dot\Theta-
G_{\dot\Theta,f_n}^*{\bf v}\}+2c_S^2\{{\bf G}_{v,f_n}^*\times\dot{\bf \Omega} +
{\bf G}_{\dot\Omega,f_n}^*\times {\bf v}\}\Bigr]\cdot{\bf n}{\rm d}^2{\bf x}.\label{eqRepcorr}
\end{eqnarray}
This is a representation of the particle velocity component $v_n({\bf x}_A,\omega)$ at ${\bf x}_A$ in $\mathbb{D}$, expressed in terms of the
wave fields ${\bf v}({\bf x},\omega)$, $\dot{\bf \Omega}({\bf x},\omega)$ and $\dot\Theta({\bf x},\omega)$ at $\partial\mathbb{D}$. 
Next, we derive a representation of the rotational motion-rate component $\dot\Omega_h({\bf x}_A,\omega)$. 
Using the subscript notation for the curl-operator,  this component is defined as
$\dot\Omega_h({\bf x}_A,\omega)=\frac{1}{2}\epsilon_{hmn}\partial_{m,A}v_n({\bf x}_A,\omega)$, where $\epsilon_{hmn}$ is the Levi-Civita symbol and where
$\partial_{m,A}$ denotes differentiation with respect to $x_{m,A}$.
Applying  the operator $\frac{1}{2}\epsilon_{hmn}\partial_{m,A}$ to both sides of equation (\ref{eqRepcorr}), interchanging the order of integration (over ${\bf x}$) and
differentiation (with respect to $x_{m,A}$), yields
\begin{eqnarray}
\dot\Omega_h({\bf x}_A,\omega)=\frac{1}{i\omega}\oint_{\partial\mathbb{D}}\rho\Bigl[c_P^2\{{\bf G}_{v,\dot\Omega_h}^*\dot\Theta-G_{\dot\Theta,\dot\Omega_h}^*{\bf v}\}
+2c_S^2\{{\bf G}_{v,\dot\Omega_h}^*\times\dot{\bf \Omega} + {\bf G}_{\dot\Omega,\dot\Omega_h}^*\times {\bf v}\}\Bigr]\cdot{\bf n}{\rm d}^2{\bf x},\label{eqRepcorrd}
\end{eqnarray}
where ${\bf G}_{\dot\Upsilon,\dot\Omega_h}({\bf x},{\bf x}_A,\omega)=\frac{1}{2}\epsilon_{hmn}\partial_{m,A}{\bf G}_{\dot\Upsilon,f_n}({\bf x},{\bf x}_A,\omega)$,
with subscript $\dot\Upsilon$ standing for $v$, $\dot\Omega$ or $\dot\Theta$. Note that here the operator $\frac{1}{2}\epsilon_{hmn}\partial_{m,A}$
transforms the force-source of the Green's function into a rotational motion source
(which for the special case of a homogeneous isotropic medium would correspond to a $S$-wave source).
A representation of the cubic dilatation-rate $\dot\Theta({\bf x}_A,\omega)$
can be derived in a similar way by applying the operator $\partial_{n,A}$  to both sides of equation (\ref{eqRepcorr}), but this is beyond the scope of this paper.

\begin{figure}
\centerline{\hspace{3.5cm}\epsfxsize=11 cm\epsfbox{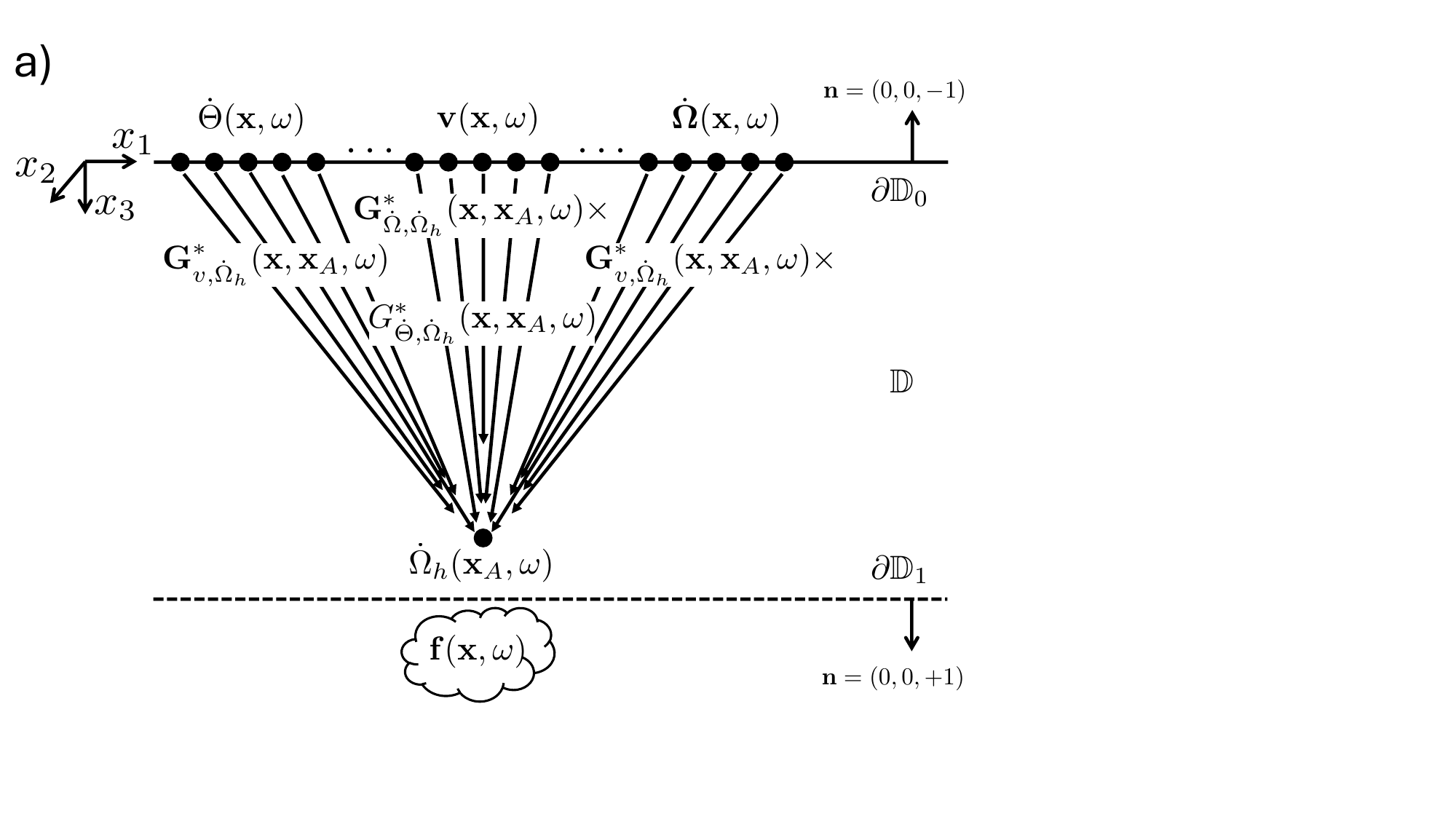}\hspace{-3cm}\epsfxsize=11 cm\epsfbox{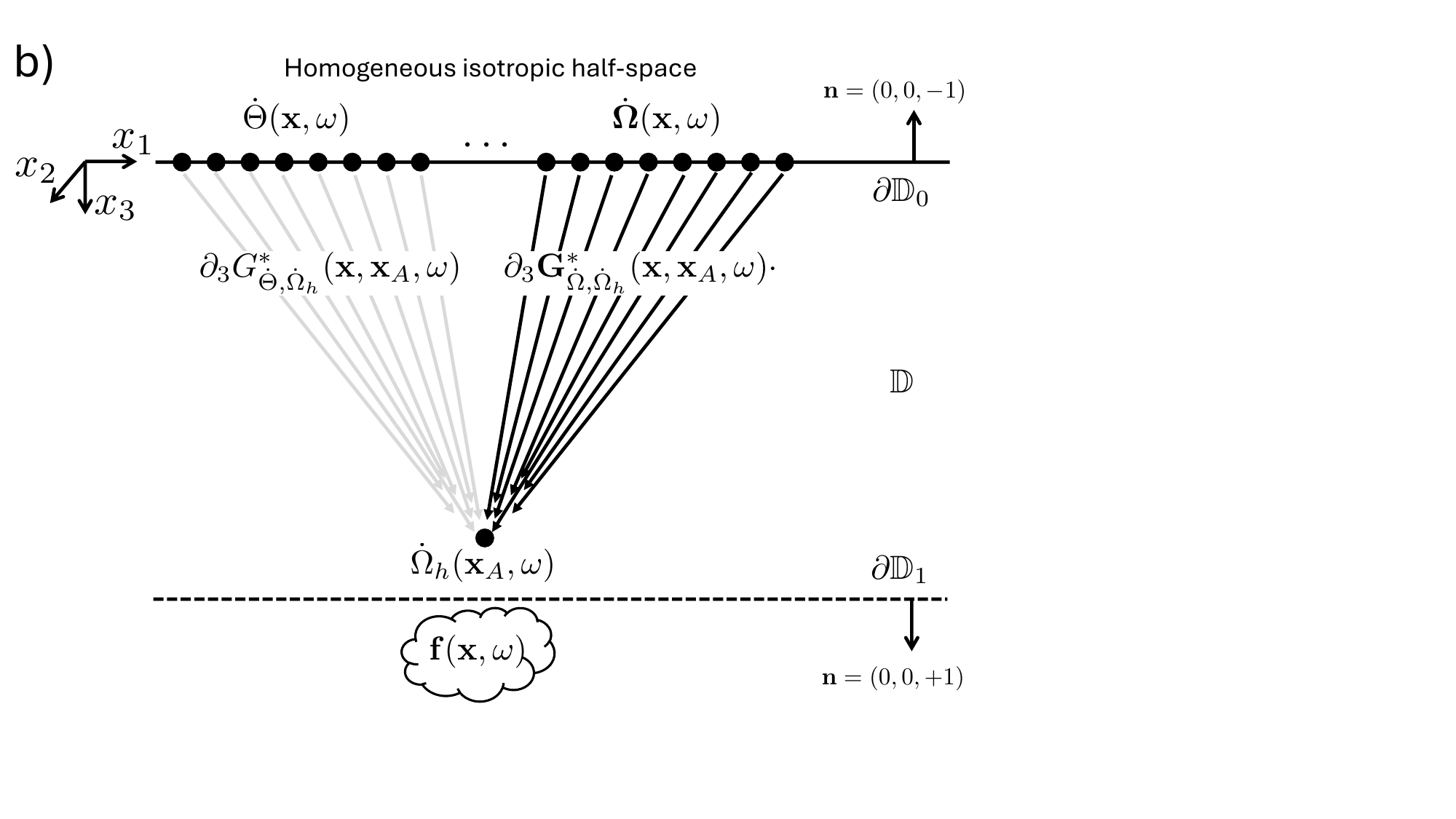}}
\vspace{-.8cm}
\caption{(a) Illustration of equation (\ref{eqRepcorrd}) (with $\partial\mathbb{D}$ replaced by $\partial\mathbb{D}_0$) 
for backpropagation. (b) Illustration of equations (\ref{eqRepcorrg}) and (\ref{eqRepcorrk}).
The amplitude of the ``converted'' Green's function $G_{\dot\Theta,\dot\Omega_h}({\bf x},{\bf x}_A,\omega)$ (light-gray) is 
one order of magnitude lower than that of the ``non-converted'' Green's function ${\bf G}_{\dot\Omega,\dot\Omega_h}({\bf x},{\bf x}_A,\omega)$.}\label{Figure1}
\end{figure}

The representations of equations (\ref{eqRepcorr}) and (\ref{eqRepcorrd}) are exact. 
In practice, however, measurements are not available on a closed boundary but, say, on a horizontal boundary $\partial\mathbb{D}_0$
(with upward pointing normal vector ${\bf n}=(0,0,-1)$). We define a second horizontal boundary $\partial\mathbb{D}_1$ 
between ${\bf x}_A$ and the source distribution ${\bf f}({\bf x},\omega)$ of the actual state, see Figure \ref{Figure1}. The  boundaries $\partial\mathbb{D}_0$ and
$\partial\mathbb{D}_1$ (together with a cylindrical boundary with a vertical axis through ${\bf x}_A$ and infinite radius) form the closed boundary $\partial\mathbb{D}$. 
Because measurements are available only on $\partial\mathbb{D}_0$, in practice we neglect the integral over $\partial\mathbb{D}_1$
and approximate equation (\ref{eqRepcorrd}) by an integral over $\partial\mathbb{D}_0$. This
is a suitable approximation for backpropagation of the wave field from $\partial\mathbb{D}_0$ to ${\bf x}_A$. 
It is illustrated in Figure \ref{Figure1}a, where the downward pointing arrows represent the complex conjugated (i.e., backpropagating) Green's functions.
These Green's functions are defined in the inhomogeneous anisotropic medium in $\mathbb{D}$,
but for simplicity they are visualized by straight rays. Since the integral over $\partial\mathbb{D}_1$ is neglected,
evanescent waves are ignored, internal multiples are erroneously handled and the 
recovered primary wave field at ${\bf x}_A$ contains small amplitude errors, proportional to the amplitudes of internal multiples \citep{Wapenaar90GP1}.
For a weakly scattering medium these approximations are acceptable and
 of the same order as those of the standard elastodynamic Kirchhoff-Helmholtz integral for  backpropagation \citep{Kuo84GEO, Hokstad2000GEO}.

Although we have achieved our goal (i.e., deriving a representation in terms of translational and rotational motion for an inhomogeneous anisotropic medium), 
equation (\ref{eqRepcorrd}) (with $\partial\mathbb{D}$ replaced by $\partial\mathbb{D}_0$)
is still rather complex. This expression simplifies significantly when the medium at and above $\partial\mathbb{D}_0$ is homogeneous and isotropic.
At and above $\partial\mathbb{D}_0$ we express the particle velocity as
 ${\bf v}=-\frac{c_P^2}{\omega^2}{\bf\nabla}\dot\Theta+\frac{2c_S^2}{\omega^2}{\bf\nabla}\times\dot{\bf \Omega}$, see Appendix B for details.
Note that $\dot\Theta$ and $\dot{\bf \Omega}$ are scaled versions of $P$- and $S$-wave potentials. We substitute 
this expression, and a similar expression for ${\bf G}_{v,\dot\Omega_h}$, into the right-hand side of equation (\ref{eqRepcorrd}).
Using the fact that the actual wave field and the Green's functions are upward propagating at $\partial\mathbb{D}_0$ 
(and hence the complex conjugated Green's functions are downward propagating at $\partial\mathbb{D}_0$), we can use one-way wave equations 
for $P$- and $S$-waves at $\partial\mathbb{D}_0$  to simplify the right-hand side of equation (\ref{eqRepcorrd}). This  yields (ignoring evanescent waves)
\begin{eqnarray}
\dot\Omega_h({\bf x}_A,\omega) \approx \frac{2}{i\omega^3}\int_{\partial\mathbb{D}_0}\rho\Bigl[c_P^4(\partial_3G_{\dot\Theta,\dot\Omega_h}^*)\dot\Theta
+4c_S^4(\partial_3{\bf G}_{\dot\Omega,\dot\Omega_h}^*)\cdot\dot{\bf \Omega} \Bigr]{\rm d}^2{\bf x},\label{eqRepcorrg}
\end{eqnarray}
see Appendix B for a detailed derivation. This expression is illustrated in Figure \ref{Figure1}b
(a similar expression was previously derived in a somewhat different way for $P$- and $S$-wave potentials  by \citet{Wapenaar90GP1}). 
Note that the Green's function $G_{\dot\Theta,\dot\Omega_h}({\bf x},{\bf x}_A,\omega)$ stands for the cubic dilatation-rate at ${\bf x}$ in response to a rotational motion source at ${\bf x}_A$.
The amplitude of this ``converted'' Green's function is one order of magnitude lower than that of the ``non-converted'' Green's function
${\bf G}_{\dot\Omega,\dot\Omega_h}({\bf x},{\bf x}_A,\omega)$ (and in a homogeneous isotropic medium it would completely vanish).
Hence, in a weakly scattering medium we can ignore the term containing $G_{\dot\Theta,\dot\Omega_h}({\bf x},{\bf x}_A,\omega)$, which leaves
\begin{eqnarray}
\dot\Omega_h({\bf x}_A,\omega) \approx \frac{8}{i\omega^3}\int_{\partial\mathbb{D}_0}
\rho c_S^4\{\partial_3{\bf G}_{\dot\Omega,\dot\Omega_h}^*({\bf x},{\bf x}_A,\omega)\}\cdot\dot{\bf \Omega}({\bf x},\omega){\rm d}^2{\bf x}.\label{eqRepcorrk}
\end{eqnarray}
This very simple Rayleigh-type integral formulates backpropagation of rotational motion-rate measurements 
$\dot{\bf \Omega}({\bf x},\omega)$ from the acquisition boundary $\partial\mathbb{D}_0$, 
through a weakly scattering inhomogeneous anistropic medium, towards sources below $\partial\mathbb{D}_0$.

\section{4\,\, Conclusions}\label{sec6}

We have derived reciprocity and representation theorems for the translational and rotational components of a seismic wave field.
Equation (\ref{eqRepcorrd}) (with $\partial\mathbb{D}$ replaced by $\partial\mathbb{D}_0$) 
 is an expression for backpropagation of measurements at the boundary $\partial\mathbb{D}_0$ 
towards real or secondary sources in the inhomogeneous anisotropic medium below $\partial\mathbb{D}_0$. 
The medium is assumed to be isotropic in a vanishingly thin shell around $\partial\mathbb{D}_0$, with $\mu$ constant along $\partial\mathbb{D}_0$.
This representation does not rely on a specific propagation direction of the wave field at $\partial\mathbb{D}_0$, hence, the medium above $\partial\mathbb{D}_0$
can also be inhomogeneous and anisotropic, or $\partial\mathbb{D}_0$ can be a free surface. 
This generality comes with complexity. When the medium above $\partial\mathbb{D}_0$ is
homogeneous and isotropic, the wave fields at $\partial\mathbb{D}_0$ propagate upward, which leads to the significantly more simple representation of equation (\ref{eqRepcorrg}) or,
when the medium below $\partial\mathbb{D}_0$ is weakly scattering, to the very simple Rayleigh-type integral of equation (\ref{eqRepcorrk}).

The derived reciprocity and representation theorems contribute to the theoretical basis for rotational seismology methodology. 
The representations of equations (\ref{eqRepcorrd}) -- (\ref{eqRepcorrk}) can be used to generate virtual rotational motion sensors inside the medium,
closer to the area of interest than the physical sensors at the surface.
Together with virtual translational motion sensors, these can be used to improve the determination of the moment tensor 
of  earthquake sources \citep{Donner2016GJI, Li2017GEO, Ichinose2021JGR} or to improve the efficiency of (local) structural imaging \citep{Bernauer2009GEO}.
Note that by rotating the configurations of Figure \ref{Figure1} by 90 degrees, similar representations  can be applied to measurements in a vertical borehole.

\section*{Acknowledgements}
The author thanks Gilles Lambar\'e and Heiner Igel for their constructive reviews, which helped to improve this paper.

\bibliographystyle{gji}

\section{Appendix A: Derivations for section 2}

The basic equations in the $({\bf x},\omega)$ domain for elastic wave fields in an inhomogeneous anisotropic medium read
\begin{eqnarray}
\partial_j\tau_{ij}+i\omega\rho v_i&=&-f_i,\label{Seq9a}\\
i\omega\tau_{ij}+c_{ijkl}\partial_lv_k&=&0.\label{Seq9b}
\end{eqnarray}
Here $v_i({\bf x},\omega)$, $\tau_{ij}({\bf x},\omega)$ and $f_i({\bf x},\omega)$ are the components of the particle velocity vector 
${\bf v}({\bf x},\omega)$, the stress tensor $\bftau({\bf x},\omega)$ and the source vector ${\bf f}({\bf x},\omega)$,
respectively; $\rho({\bf x})$ is the mass density and $c_{ijkl}({\bf x})$ the stiffness.  Einstein's summation convention applies to repeated lower-case subscripts. 
Consider a spatial domain $\mathbb{D}$ enclosed by boundary $\partial\mathbb{D}$.
For this configuration, the Betti-Rayleigh reciprocity theorem reads
%
\begin{eqnarray}
&&\hspace{-.8cm}\oint_{\partial\mathbb{D}}\{v_{i,A}\tau_{ij,B}-v_{i,B}\tau_{ij,A}\}n_j{\rm d}^2{\bf x}=-\int_\mathbb{D}\{v_{i,A}f_{i,B}-v_{i,B}f_{i,A}\}{\rm d}^3{\bf x},\label{eqS9}
\end{eqnarray}
where upper-case subscripts $A$ and $B$ denote two independent states and 
$n_j$ stands for the components of the outward pointing normal vector ${\bf n}$ on $\partial\mathbb{D}$.

We recast equation (\ref{eqS9}) into a form that contains the rotational motion-rate vector $\frac{1}{2}{\bf\nabla}\times{\bf v}$ in states $A$ and $B$.
To this end we assume that the medium is isotropic in a vanishingly thin shell around the boundary $\partial\mathbb{D}$, hence
\begin{eqnarray}
c_{ijkl}=\lambda\delta_{ij}\delta_{kl}+\mu(\delta_{ik}\delta_{jl}+\delta_{il}\delta_{jk}),\label{Seq100}
\end{eqnarray}
where $\lambda({\bf x})$ and $\mu({\bf x})$ are the Lam\'e parameters in the vanishingly thin shell around $\partial\mathbb{D}$.
Substituting equation (\ref{Seq100}) into equation (\ref{Seq9b}) gives
\begin{eqnarray}
\tau_{ij}=-\frac{1}{i\omega}\Bigl(\lambda\delta_{ij}\partial_kv_k + \mu(\partial_jv_i+\partial_iv_j)\Bigr).\label{eqS10}
\end{eqnarray}
Next, substituting equation (\ref{eqS10}) into equation (\ref{eqS9}) yields
\begin{eqnarray}
&&\hspace{-.8cm}\oint_{\partial\mathbb{D}}\Bigl[\lambda\{v_{j,A}\partial_kv_{k,B} - v_{j,B}\partial_kv_{k,A}\}
+\mu\{v_{i,A}(\partial_jv_{i,B}+\partial_iv_{j,B}) - v_{i,B}(\partial_jv_{i,A}+\partial_iv_{j,A})\}\Bigr]n_j{\rm d}^2{\bf x}\nonumber\\
&&\hspace{.43cm}=i\omega\int_\mathbb{D}\{v_{i,A}f_{i,B}-v_{i,B}f_{i,A}\}{\rm d}^3{\bf x}.\label{eqS11}
\end{eqnarray}
Using
\begin{eqnarray}
v_{j,A}\partial_kv_{k,B}&=&\{{\bf v}_A{\bf\nabla}\cdot{\bf v}_B\}_j,\\
v_{i,A}(\partial_jv_{i,B}+\partial_iv_{j,B})&=&\{2({\bf v}_A\cdot{\bf\nabla}){\bf v}_B+{\bf v}_A\times{\bf\nabla}\times{\bf v}_B\}_j,\nonumber\\
v_{i,A}f_{i,B}&=&{\bf v}_A\cdot{\bf f}_B,
\end{eqnarray}
and similar expressions with $A$ and $B$ interchanged, equation (\ref{eqS11}) can be rewritten as
\begin{eqnarray}
&&\hspace{-.83cm}\oint_{\partial\mathbb{D}}\Bigl[\lambda\{{\bf v}_A{\bf\nabla}\cdot{\bf v}_B-{\bf v}_B{\bf\nabla}\cdot{\bf v}_A\}
+2\mu\{({\bf v}_A\cdot{\bf\nabla}){\bf v}_B - ({\bf v}_B\cdot{\bf\nabla}){\bf v}_A\}\nonumber\\
&&\hspace{.43cm}+\mu\{{\bf v}_A\times{\bf\nabla}\times{\bf v}_B - {\bf v}_B\times{\bf\nabla}\times{\bf v}_A\}\Bigr]\cdot{\bf n}{\rm d}^2{\bf x}
=i\omega\int_\mathbb{D}\{{\bf v}_A\cdot{\bf f}_B - {\bf v}_B\cdot{\bf f}_A\}{\rm d}^3{\bf x}.\label{eqS14}
\end{eqnarray}
This is equation (3) in the main text.

Next, we aim to combine the first two terms on the left-hand side into a single term.
Assuming $\mu$ is constant along the boundary $\partial\mathbb{D}$, we can take it outside the integral and
use the theorem of Gauss to reorganize the second term into the form of the first term, according to
\begin{eqnarray}
&&\hspace{-.43cm}2\mu\oint_{\partial\mathbb{D}}\{({\bf v}_A\cdot{\bf\nabla}){\bf v}_B - ({\bf v}_B\cdot{\bf\nabla}){\bf v}_A\}\cdot{\bf n}{\rm d}^2{\bf x}
=2\mu\int_\mathbb{D}{\bf\nabla}\cdot\{({\bf v}_A\cdot{\bf\nabla}){\bf v}_B - ({\bf v}_B\cdot{\bf\nabla}){\bf v}_A\}{\rm d}^3{\bf x}\nonumber\\
&&\hspace{-.43cm}=2\mu\int_\mathbb{D}{\bf\nabla}\cdot\{{\bf v}_A{\bf\nabla}\cdot{\bf v}_B-{\bf v}_B{\bf\nabla}\cdot{\bf v}_A\}{\rm d}^3{\bf x}
=2\mu\oint_{\partial\mathbb{D}}\{{\bf v}_A{\bf\nabla}\cdot{\bf v}_B-{\bf v}_B{\bf\nabla}\cdot{\bf v}_A\}\cdot{\bf n}{\rm d}^2{\bf x}.\label{eqS15}
\end{eqnarray}
%
%
Using equation (\ref{eqS15}) in equation (\ref{eqS14}) we obtain
%
\begin{eqnarray}
&&\hspace{-.8cm}\oint_{\partial\mathbb{D}}\rho\Bigl[c_P^2\{{\bf v}_A{\bf\nabla}\cdot{\bf v}_B - {\bf v}_B{\bf\nabla}\cdot{\bf v}_A\}
+c_S^2\{{\bf v}_A\times{\bf\nabla}\times{\bf v}_B - {\bf v}_B\times{\bf\nabla}\times{\bf v}_A\}\Bigr]\cdot{\bf n}{\rm d}^2{\bf x}\nonumber\\
&&\hspace{.43cm}=i\omega\int_\mathbb{D}\{{\bf v}_A\cdot{\bf f}_B - {\bf v}_B\cdot{\bf f}_A\}{\rm d}^3{\bf x},\label{eqS20000}
\end{eqnarray}
with $P$- and $S$-propagation velocities $c_P = \sqrt{(\lambda+2\mu)/\rho}$ and $c_S = \sqrt{\mu/\rho}$ at $\partial\mathbb{D}$.
This is equation (1) in the main text, but for an inhomogeneous anisotropic medium in $\mathbb{D}$;
only at the boundary $\partial\mathbb{D}$ the medium is assumed to be isotropic, with $\mu$ constant along $\partial\mathbb{D}$.
Using
\begin{eqnarray}
\dot\Theta&=&{\bf\nabla}\cdot{\bf v},\label{eqS7}\\
\dot{\bf \Omega}&=&\frac{1}{2}{\bf\nabla}\times{\bf v},\label{eqS5}
\end{eqnarray}
%
equation (\ref{eqS20000}) can be rewritten as
\begin{eqnarray}
&&\hspace{-.8cm}\oint_{\partial\mathbb{D}}\rho\Bigl[c_P^2\{{\bf v}_A\dot\Theta_B-{\bf v}_B\dot\Theta_A\}
+2c_S^2\{{\bf v}_A\times\dot{\bf \Omega}_B - {\bf v}_B\times\dot{\bf \Omega}_A\}\Bigr]\cdot{\bf n}{\rm d}^2{\bf x}
=i\omega\int_\mathbb{D}\{{\bf v}_A\cdot{\bf f}_B - {\bf v}_B\cdot{\bf f}_A\}{\rm d}^3{\bf x}.\nonumber\\
&&\label{eqS16b}
\end{eqnarray}

We derive a similar expression with complex-conjugated fields in one of the states. Assuming the medium is lossless (i.e., assuming $\rho$ and $c_{ijkl}$ are real-valued),
complex conjugation of equations (\ref{Seq9a}) and (\ref{Seq9b}) yields 
\begin{eqnarray}
-\partial_j\tau_{ij}^*+i\omega\rho v_i^*&=&f_i^*,\label{Seq9aconj}\\
-i\omega\tau_{ij}^*+c_{ijkl}\partial_lv_k^*&=&0,\label{Seq9bconj}
\end{eqnarray}
where the asterisk denotes complex conjugation.
Hence, since $v_i^*$, $-\tau_{ij}^*$ and $-f_i^*$ obey the same equations as $v_i$, $\tau_{ij}$ and $f_i$, the Betti-Rayleigh reciprocity theorem of equation (\ref{eqS9}) can be 
modified into
\begin{eqnarray}
&&\hspace{-.8cm}\oint_{\partial\mathbb{D}}\{v_{i,A}^*\tau_{ij,B}+v_{i,B}\tau_{ij,A}^*\}n_j{\rm d}^2{\bf x}=-\int_\mathbb{D}\{v_{i,A}^*f_{i,B}+v_{i,B}f_{i,A}^*\}{\rm d}^3{\bf x}.\label{eqS9conj}
\end{eqnarray}
Following the same derivation as above, we obtain
\begin{eqnarray}
&&\hspace{-.8cm}\oint_{\partial\mathbb{D}}\rho\Bigl[c_P^2\{{\bf v}_A^*\dot\Theta_B-{\bf v}_B\dot\Theta_A^*\}
+2c_S^2\{{\bf v}_A^*\times\dot{\bf \Omega}_B - {\bf v}_B\times\dot{\bf \Omega}_A^*\}\Bigr]\cdot{\bf n}{\rm d}^2{\bf x}
=i\omega\int_\mathbb{D}\{{\bf v}_A^*\cdot{\bf f}_B + {\bf v}_B\cdot{\bf f}_A^*\}{\rm d}^3{\bf x}.\nonumber\\
&&\label{eqS16bconj}
\end{eqnarray}
This is equation (4) in the main text.


%

\section{Appendix B: Derivations for section 3}

Consider the  boundary integral of equation (6) in the main text, with $\partial\mathbb{D}$ replaced by the horizontal boundary $\partial\mathbb{D}_0$, i.e.,
\begin{eqnarray}
\int_{\partial\mathbb{D}_0}\rho\Bigl[c_P^2\{{\bf G}_{v,\dot\Omega_h}^*\dot\Theta-
G_{\dot\Theta,\dot\Omega_h}^*{\bf v}\}
+2c_S^2\{{\bf G}_{v,\dot\Omega_h}^*\times\dot{\bf \Omega} +
{\bf G}_{\dot\Omega,\dot\Omega_h}^*\times {\bf v}\}\Bigr]\cdot{\bf n}{\rm d}^2{\bf x},\label{eqSRepcorrf}
\end{eqnarray}
with ${\bf n}=(0,0,-1)$ at $\partial\mathbb{D}_0$.
Our aim is to reorganize this integral
for the situation that the medium at and above $\partial\mathbb{D}_0$ is homogeneous, isotropic and source-free. 
At and above $\partial\mathbb{D}_0$,  the velocity vector ${\bf v}({\bf x},\omega)$ obeys the following wave equation
\begin{eqnarray}
c_P^2{\bf\nabla}({\bf\nabla}\cdot{\bf v})-c_S^2{\bf\nabla}\times{\bf\nabla}\times{\bf v}+\omega^2{\bf v}= {\bf 0}.\label{eqS1}
\end{eqnarray}
%
%
%
For the homogeneous, isotropic medium at and above $\partial\mathbb{D}_0$ we write
 \begin{eqnarray}
 {\bf v}=a{\bf\nabla}\dot\Theta+b{\bf\nabla}\times\dot{\bf \Omega},\label{eqS24}
 \end{eqnarray}
where $a$ and $b$ still need to be determined.
Comparing this expression with ${\bf v}={\bf\nabla}\Phi+{\bf\nabla}\times{\bf \Psi}$, where $\Phi$ and ${\bf \Psi}$ are the $P$- and $S$-wave potentials,
it follows that $\dot\Theta$ and $\dot{\bf \Omega}$ are proportional to the $P$- and $S$-wave potentials.
Note that
%
\begin{eqnarray}
{\bf\nabla}\cdot{\bf v}&=&a{\bf\nabla}^2\dot\Theta,\label{eqS25}\\
{\bf\nabla}\times{\bf v}&=&b{\bf\nabla}\times{\bf\nabla}\times\dot{\bf \Omega}.\label{eqS26}
\end{eqnarray}
Substitution of equations (\ref{eqS24}) -- (\ref{eqS26}) into equation (\ref{eqS1}) gives
\begin{eqnarray}
a c_P^2{\bf\nabla}\Bigl({\bf\nabla}^2\dot\Theta+\frac{\omega^2}{c_P^2}\dot\Theta\Bigr) + 
b c_S^2{\bf\nabla}\times\Bigl(-{\bf\nabla}\times{\bf\nabla}\times\dot{\bf \Omega}+\frac{\omega^2}{c_S^2}\dot{\bf \Omega}\Bigr)=0.\label{eqS27}
\end{eqnarray}
The decomposition of this equation into independent equations for $\dot\Theta$ and $\dot{\bf \Omega}$ is not unique. A convenient choice is
\begin{eqnarray}
{\bf\nabla}^2\dot\Theta+\frac{\omega^2}{c_P^2}\dot\Theta&=&0,\label{eqS28}\\
-{\bf\nabla}\times{\bf\nabla}\times\dot{\bf \Omega}+\frac{\omega^2}{c_S^2}\dot{\bf \Omega}&=&0.\label{eqS29}
\end{eqnarray}
Equation (\ref{eqS29})  implies ${\bf\nabla}\cdot\dot{\bf \Omega}=0$ (for $\omega\ne 0$). Using the fundamental property
\begin{eqnarray}
-{\bf\nabla}\times{\bf\nabla}\times\dot{\bf \Omega}+{\bf\nabla}({\bf\nabla}\cdot\dot{\bf \Omega})={\bf\nabla}^2\dot{\bf \Omega},
\end{eqnarray}
equation (\ref{eqS29}) can be rewritten as
\begin{eqnarray}
{\bf\nabla}^2\dot{\bf \Omega}+\frac{\omega^2}{c_S^2}\dot{\bf \Omega}={\bf 0},\quad \mbox{with}\quad {\bf\nabla}\cdot\dot{\bf \Omega}=0.\label{eqS31}
\end{eqnarray}
From equations (\ref{eqS7}), (\ref{eqS25}) and (\ref{eqS28}) we obtain
\begin{eqnarray}
a=-\frac{c_P^2}{\omega^2}.
\end{eqnarray}
Similarly, from equations (\ref{eqS5}), (\ref{eqS26}) and (\ref{eqS29}) we obtain
\begin{eqnarray}
b=\frac{2c_S^2}{\omega^2}.
\end{eqnarray}
Substitution into equation (\ref{eqS24}) gives
 \begin{eqnarray}
 {\bf v}=-\frac{c_P^2}{\omega^2}{\bf\nabla}\dot\Theta+\frac{2c_S^2}{\omega^2}{\bf\nabla}\times\dot{\bf \Omega}.\label{eqS34}
 \end{eqnarray}
We express the Green's velocity vector in a similar way, according to
 \begin{eqnarray}
 {\bf G}_{v,\dot\Omega_h}=-\frac{c_P^2}{\omega^2}{\bf\nabla}G_{\dot\Theta,\dot\Omega_h}+\frac{2c_S^2}{\omega^2}{\bf\nabla}\times{\bf G}_{\dot\Omega,\dot\Omega_h}.\label{eqS35}
 \end{eqnarray}
%
%
Substitution of equations (\ref{eqS34}) and (\ref{eqS35}) into the boundary integral of equation (\ref{eqSRepcorrf}) yields
\begin{eqnarray}
&&\hspace{-.6cm}\frac{\rho}{\omega^2}\int_{\partial\mathbb{D}_0}\Bigl[c_P^2\Bigl(-(c_P^2{\bf\nabla}G_{\dot\Theta,\dot\Omega_h}-2c_S^2{\bf\nabla}\times{\bf G}_{\dot\Omega,\dot\Omega_h})^*\dot\Theta
+G_{\dot\Theta,\dot\Omega_h}^*(c_P^2{\bf\nabla}\dot\Theta-2c_S^2{\bf\nabla}\times\dot{\bf \Omega})\Bigr)\nonumber\\
&&\hspace{-.3cm}-2c_S^2\Bigl((c_P^2{\bf\nabla}G_{\dot\Theta,\dot\Omega_h}-2c_S^2{\bf\nabla}\times{\bf G}_{\dot\Omega,\dot\Omega_h})^*\times\dot{\bf \Omega} 
+{\bf G}_{\dot\Omega,\dot\Omega_h}^*\times(c_P^2{\bf\nabla}\dot\Theta-2c_S^2{\bf\nabla}\times\dot{\bf \Omega})\Bigr) \Bigr]\cdot{\bf n}{\rm d}^2{\bf x}.\label{eqS35b}
\end{eqnarray}
%
The medium at and above $\partial\mathbb{D}_0$ is homogeneous, isotropic and source-free.
The source of the Green's function (at ${\bf x}_A$) and the source distribution ${\bf f}({\bf x},\omega)$ of the actual field are below the boundary $\partial\mathbb{D}_0$,
hence, the Green's function and the actual field propagate upward at $\partial\mathbb{D}_0$. In the following we make explicitly use of this.
%
To this end, we first rewrite the boundary integral in equation (\ref{eqS35b}) as
\begin{eqnarray}
\int_{\partial\mathbb{D}_0}\Bigl[\cdots\Bigr]\cdot{\bf n}{\rm d}^2{\bf x}=\int_{\mathbb{R}^2}\Bigl[\cdots\Bigr]_{x_{3,0}}\cdot{\bf n}{\rm d}^2{\bf x}_{\rm H},\label{eqS18}
\end{eqnarray}
where  $\mathbb{R}$ is the set of real numbers, $x_{3,0}$ is the depth of boundary $\partial\mathbb{D}_0$ and ${\bf x}_{\rm H}$ denotes the horizontal coordinate vector, defined as ${\bf x}_{\rm H}=(x_1,x_2)$.
We define the 2D spatial Fourier transform of a quantity $A({\bf x}_{\rm H},x_{3,0},\omega)$ as
\begin{eqnarray}
\tilde A({\bf k}_{\rm H},x_{3,0},\omega)=\int_{\mathbb{R}^2}A({\bf x}_{\rm H},x_{3,0},\omega)\exp\{-i {\bf k}_{\rm H}\cdot{\bf x}_{\rm H}\}{\rm d}^2{\bf x}_{\rm H},
\end{eqnarray}
with ${\bf k}_{\rm H}$ denoting the horizontal wave vector, according to ${\bf k}_{\rm H}=(k_1,k_2)$.
Using equation (\ref{eqS18}) and Parseval's theorem
\begin{eqnarray}
\int_{\mathbb{R}^2}A^*({\bf x}_{\rm H},x_{3,0},\omega)B({\bf x}_{\rm H},x_{3,0},\omega){\rm d}^2{\bf x}_{\rm H}
=\frac{1}{4\pi^2}\int_{\mathbb{R}^2}\tilde A^*({\bf k}_{\rm H},x_{3,0},\omega)\tilde B({\bf k}_{\rm H},x_{3,0},\omega){\rm d}^2{\bf k}_{\rm H},\label{eqSParseval2}
\end{eqnarray}
we rewrite the expression of equation (\ref{eqS35b}) as
\begin{eqnarray}
&&\hspace{-.8cm}\frac{\rho}{4\pi^2\omega^2}\int_{\mathbb{R}^2}\Bigl[c_P^2\Bigl(-(c_P^2\tilde{\bf\nabla}\tilde G_{\dot\Theta,\dot\Omega_h}-2c_S^2\tilde{\bf\nabla}\times{\tilde{\bf G}}_{\dot\Omega,\dot\Omega_h})^*\tilde{\dot\Theta}
+\tilde G_{\dot\Theta,\dot\Omega_h}^*(c_P^2\tilde{\bf\nabla}\tilde{\dot\Theta}-2c_S^2\tilde{\bf\nabla}\times{\tilde{\dot{\bf \Omega}}})\Bigr)-\nonumber\\
&&\hspace{-.8cm}2c_S^2\Bigl((c_P^2\tilde{\bf\nabla}\tilde G_{\dot\Theta,\dot\Omega_h}-2c_S^2\tilde{\bf\nabla}\times{\tilde{\bf G}}_{\dot\Omega,\dot\Omega_h})^*\times{\tilde{\dot{\bf \Omega}}} 
+\tilde{\bf G}_{\dot\Omega,\dot\Omega_h}^*\times(c_P^2\tilde{\bf\nabla}\tilde{\dot\Theta}-2c_S^2\tilde{\bf\nabla}\times{\tilde{\dot{\bf \Omega}}})\Bigr) \Bigr]_{x_{3,0}}\cdot{\bf n}{\rm d}^2{\bf k}_{\rm H},\label{eqS36b}
\end{eqnarray}
with
\begin{eqnarray}
\tilde{\bf\nabla}=\begin{pmatrix}ik_1\\ik_2\\\partial_3\end{pmatrix}.\label{eqS22}
\end{eqnarray}
Equations (\ref{eqS28}) and (\ref{eqS31}) read in the spatial Fourier domain
\begin{eqnarray}
\partial_3^2\tilde{\dot\Theta}&=&-\biggl(\frac{\omega^2}{c_P^2}-{\bf k}_{\rm H}\cdot{\bf k}_{\rm H}\biggr)\tilde{\dot\Theta},\label{eqS28b}\\
\partial_3^2{\tilde{\dot{\bf \Omega}}}&=&-\biggl(\frac{\omega^2}{c_P^2}-{\bf k}_{\rm H}\cdot{\bf k}_{\rm H}\biggr){\tilde{\dot{\bf \Omega}}},
\quad \mbox{with}\quad ik_1\tilde{\dot\Omega}_1+ik_2\tilde{\dot\Omega}_2+\partial_3\tilde{\dot\Omega}_3=0.\label{eqS31b}
\end{eqnarray}
Given that $\tilde{\dot\Theta}$ and ${\tilde{\dot{\bf \Omega}}}$ are propagating upward at $x_3=x_{3,0}$, we can use the following one-way wave equations
\begin{eqnarray}
\partial_3\tilde{\dot\Theta}&=& -ik_{3,P}\tilde{\dot\Theta},\label{eqS41}\\
\partial_3\tilde{\dot{{\bf \Omega}}}&=& -ik_{3,S}\tilde{\dot{{\bf \Omega}}},\quad \mbox{with}\quad ik_1\tilde{\dot\Omega}_1+ik_2\tilde{\dot\Omega}_2=ik_{3,S}\tilde{\dot\Omega}_3,\label{eqS42}
\end{eqnarray}
where
\begin{eqnarray}
k_{3,P}&=&\begin{cases}
\sqrt{\frac{\omega^2}{c_P^2}-{\bf k}_{\rm H}\cdot{\bf k}_{\rm H}},\quad&\mbox{for}\quad{\bf k}_{\rm H}\cdot{\bf k}_{\rm H}\le\frac{\omega^2}{c_P^2},\\
i\sqrt{{\bf k}_{\rm H}\cdot{\bf k}_{\rm H}-\frac{\omega^2}{c_P^2}},\quad&\mbox{for}\quad{\bf k}_{\rm H}\cdot{\bf k}_{\rm H}>\frac{\omega^2}{c_P^2},
\end{cases}\label{eqS27h}\\
k_{3,S}&=&\begin{cases}
\sqrt{\frac{\omega^2}{c_S^2}-{\bf k}_{\rm H}\cdot{\bf k}_{\rm H}},\quad&\mbox{for}\quad{\bf k}_{\rm H}\cdot{\bf k}_{\rm H}\le\frac{\omega^2}{c_S^2},\\
i\sqrt{{\bf k}_{\rm H}\cdot{\bf k}_{\rm H}-\frac{\omega^2}{c_S^2}},\quad&\mbox{for}\quad{\bf k}_{\rm H}\cdot{\bf k}_{\rm H}>\frac{\omega^2}{c_S^2}.
\end{cases}\label{eqS28h}
\end{eqnarray}
Note that for ${\bf k}_{\rm H}\cdot{\bf k}_{\rm H}\le{\omega^2}/{c_P^2}$ and for ${\bf k}_{\rm H}\cdot{\bf k}_{\rm H}>{\omega^2}/{c_P^2}$, equation (\ref{eqS41}) 
describes upward propagating and upward decaying evanescent  $P$-waves, respectively. Similarly, 
for ${\bf k}_{\rm H}\cdot{\bf k}_{\rm H}\le{\omega^2}/{c_S^2}$ and for ${\bf k}_{\rm H}\cdot{\bf k}_{\rm H}>{\omega^2}/{c_S^2}$, equation (\ref{eqS42}) 
describes upward propagating and upward decaying evanescent  $S$-waves, respectively. 
Analogous to equations (\ref{eqS41}) and (\ref{eqS42}) we have
\begin{eqnarray}
(\partial_3\tilde G_{\dot\Theta,\dot\Omega_h})^*&=& ik_{3,P}^*\tilde G_{\dot\Theta,\dot\Omega_h}^*,\label{eqS41h}\\
(\partial_3{\tilde{\bf G}}_{\dot\Omega,\dot\Omega_h})^*&=& ik_{3,S}^*{\tilde{\bf G}}_{\dot\Omega,\dot\Omega_h}^*,\quad \mbox{with}\quad
- ik_1\tilde G_{\dot\Omega_1,\dot\Omega_h}^*-ik_2\tilde G_{\dot\Omega_2,\dot\Omega_h}^*=-ik_{3,S}^*\tilde G_{\dot\Omega_3,\dot\Omega_h}^*\label{eqS42h}
\end{eqnarray}
($\tilde G_{\dot\Omega_k,\dot\Omega_h}$ being the $k$-component of $\tilde {\bf G}_{\dot\Omega,\dot\Omega_h}$),
where, according to equations (\ref{eqS27h}) and (\ref{eqS28h}),
\begin{eqnarray}
k_{3,P}^*&=&\begin{cases}
k_{3,P},\quad&\mbox{for}\quad{\bf k}_{\rm H}\cdot{\bf k}_{\rm H}\le\frac{\omega^2}{c_P^2},\\
-k_{3,P},\quad&\mbox{for}\quad{\bf k}_{\rm H}\cdot{\bf k}_{\rm H}>\frac{\omega^2}{c_P^2},\label{eqS36g}
\end{cases}\\
k_{3,S}^*&=&\begin{cases}
k_{3,S},\quad&\mbox{for}\quad{\bf k}_{\rm H}\cdot{\bf k}_{\rm H}\le\frac{\omega^2}{c_S^2},\\
-k_{3,S},\quad&\mbox{for}\quad{\bf k}_{\rm H}\cdot{\bf k}_{\rm H}>\frac{\omega^2}{c_S^2}.\label{eqS36h}
\end{cases}
\end{eqnarray}
Using equations (\ref{eqS22}) and (\ref{eqS41}) -- (\ref{eqS42h}),  we can reorganize the expression of equation (\ref{eqS36b}) as
\begin{eqnarray}
&&\hspace{-.8cm}-\frac{\rho}{4\pi^2\omega^2}\int_{\mathbb{R}^2}\Bigl[c_P^4\Bigl(-(\partial_3\tilde G_{\dot\Theta,\dot\Omega_h})^*\tilde{\dot\Theta}+\tilde G_{\dot\Theta,\dot\Omega_h}^*\partial_3\tilde{\dot\Theta}\Bigr)\nonumber\\
&&\hspace{1.2cm}+2c_P^2c_S^2\Bigl((ik_1\tilde G_{\dot\Omega_2,\dot\Omega_h}-ik_2\tilde G_{\dot\Omega_1,\dot\Omega_h})^*\tilde{\dot\Theta}-
\tilde G_{\dot\Theta,\dot\Omega_h}^*(ik_1\tilde{\dot\Omega}_2-ik_2\tilde{\dot\Omega}_1)\nonumber\\
&&\hspace{2.4cm}-(ik_1\tilde G_{\dot\Theta,\dot\Omega_h})^*\tilde{\dot\Omega}_2+(ik_2\tilde G_{\dot\Theta,\dot\Omega_h})^*\tilde{\dot\Omega}_1
-\tilde G_{\dot\Omega_1,\dot\Omega_h}^*ik_2\tilde{\dot\Theta}+\tilde G_{\dot\Omega_2,\dot\Omega_h}^*ik_1\tilde{\dot\Theta}\Bigr)\nonumber\\
&&\hspace{1.2cm}+4c_S^4\Bigl((ik_2\tilde G_{\dot\Omega_3,\dot\Omega_h}-\partial_3\tilde G_{\dot\Omega_2,\dot\Omega_h})^*\tilde{\dot\Omega}_2
-(\partial_3\tilde G_{\dot\Omega_1,\dot\Omega_h}-ik_1\tilde G_{\dot\Omega_3,\dot\Omega_h})^*\tilde{\dot\Omega}_1\nonumber\\
&&\hspace{2.4cm}+\tilde G_{\dot\Omega_1,\dot\Omega_h}^*(\partial_3\tilde{\dot\Omega}_1-ik_1\tilde{\dot\Omega}_3)
-\tilde G_{\dot\Omega_2,\dot\Omega_h}^*(ik_2\tilde{\dot\Omega}_3-\partial_3\tilde{\dot\Omega}_1)\Bigr) \Bigr]_{x_{3,0}}{\rm d}^2{\bf k}_{\rm H}\nonumber\\
&&\hspace{-.8cm}=\frac{\rho}{4\pi^2\omega^2}\int_{\mathbb{R}^2}\Bigl[c_P^4(ik_{3,P}+ik_{3,P}^*)\tilde G_{\dot\Theta,\dot\Omega_h}^*\tilde{\dot\Theta}
+4c_S^4\Bigl((ik_{3,S}+ik_{3,S}^*)(\tilde G_{\dot\Omega_1,\dot\Omega_h}^*\tilde{\dot\Omega}_1+\tilde G_{\dot\Omega_2,\dot\Omega_h}^*\tilde{\dot\Omega}_2)\nonumber\\
&&\hspace{1.2cm}+\tilde G_{\dot\Omega_3,\dot\Omega_h}^*(ik_1\tilde{\dot\Omega}_1+ik_2\tilde{\dot\Omega}_2)+
(ik_1\tilde G_{\dot\Omega_1,\dot\Omega_h}^*+ik_2\tilde G_{\dot\Omega_2,\dot\Omega_h}^*)\tilde{\dot\Omega}_3\Bigr) \Bigr]_{x_{3,0}}{\rm d}^2{\bf k}_{\rm H}\nonumber\\
&&\hspace{-.8cm}=\frac{\rho}{4\pi^2\omega^2}\int_{\mathbb{R}^2}\Bigl[c_P^4(ik_{3,P}+ik_{3,P}^*)\tilde G_{\dot\Theta,\dot\Omega_h}^*\tilde{\dot\Theta}\nonumber\\
&&\hspace{1.2cm}+4c_S^4\Bigl((ik_{3,S}+ik_{3,S}^*)(\tilde G_{\dot\Omega_1,\dot\Omega_h}^*\tilde{\dot\Omega}_1+\tilde G_{\dot\Omega_2,\dot\Omega_h}^*\tilde{\dot\Omega}_2
+\tilde G_{\dot\Omega_3,\dot\Omega_h}^*\tilde{\dot\Omega}_3)\Bigr) \Bigr]_{x_{3,0}}{\rm d}^2{\bf k}_{\rm H}.\label{eqSall}
\end{eqnarray}
From equations (\ref{eqS36g}) and (\ref{eqS36h}) it follows that the integrand of equation (\ref{eqSall}) vanishes for the evanescent wave regimes. 
By restricting the integrals to the propagating wave regimes, we obtain (using again equations (\ref{eqS41h}) and (\ref{eqS42h}))
%
%
%
\begin{eqnarray}
&&\hspace{-.8cm}\frac{2\rho}{4\pi^2\omega^2}\Biggl[\int_{{\bf k}_{\rm H}\cdot{\bf k}_{\rm H}\le\frac{\omega^2}{c_P^2}} \bigl[c_P^4(\partial_3\tilde G_{\dot\Theta,\dot\Omega_h}^*)\tilde{\dot\Theta}\bigr]_{x_{3,0}}{\rm d}^2{\bf k}_{\rm H}
+\int_{{\bf k}_{\rm H}\cdot{\bf k}_{\rm H}\le\frac{\omega^2}{c_S^2}} \bigl[4c_S^4(\partial_3{\tilde{\bf G}}_{\dot\Omega,\dot\Omega_h}^*)\cdot{\tilde{\dot{\bf \Omega}}}\bigr]_{x_{3,0}}{\rm d}^2{\bf k}_{\rm H}\Biggr].\label{eqS99}
\end{eqnarray}
%
Up to this point we have made no approximations, hence,
the expression in equation (\ref{eqS99}) is identical to that in equation (\ref{eqSRepcorrf}). Next, we want to apply Parseval's theorem again to obtain a space-domain integral.
To this end, we first extend the integration intervals in equation (\ref{eqS99}) to $\mathbb{R}^2$, according to
\begin{eqnarray}
&&\hspace{-.6cm}\frac{2\rho}{4\pi^2\omega^2}\int_{\mathbb{R}^2} \Bigl[c_P^4(\partial_3\tilde G_{\dot\Theta,\dot\Omega_h}^*)\tilde{\dot\Theta}+
4c_S^4(\partial_3{\tilde{\bf G}}_{\dot\Omega,\dot\Omega_h}^*)\cdot{\tilde{\dot{\bf \Omega}}}\Bigr]_{x_{3,0}}{\rm d}^2{\bf k}_{\rm H}.\label{eqS100}
\end{eqnarray}
This is a reasonable approximation, since the wave fields under the extended integral are negligible in the evanescent wave regimes. 
%
%
 Using Parseval's theorem (equation (\ref{eqSParseval2})) and equation (\ref{eqS18}) 
(without the inner product with ${\bf n}$), we finally obtain
\begin{eqnarray}
 &&\hspace{-.6cm}\frac{2\rho}{\omega^2}\int_{\partial\mathbb{D}_0}\Bigl[c_P^4(\partial_3G_{\dot\Theta,\dot\Omega_h}^*)\dot\Theta
+4c_S^4(\partial_3{\bf G}_{\dot\Omega,\dot\Omega_h}^*)\cdot\dot{\bf \Omega} \Bigr]{\rm d}^2{\bf x}.\label{eqSRepcorrg}
\end{eqnarray}
Substituting this for the  boundary integral of equation (6) in the main text
(with $\partial\mathbb{D}$ replaced by $\partial\mathbb{D}_0$), we obtain equation (7) in the main text.

\end{spacing}
\end{document}